\documentclass[aip,apl,twocolumn,10pt,superscriptaddress]{revtex4-1}

\usepackage{graphicx}
\usepackage{amssymb}
\usepackage{wasysym}
\usepackage{natbib}
\usepackage{color}
\usepackage[dvipdfmx,colorlinks=true,citecolor=blue,linkcolor=blue,pdfhighlight =/O]{hyperref}

\renewcommand{\refname}{} 
\begin{document}

\title{\textcolor{blue}{\Large Absence of boron aggregates in superconducting silicon confirmed by atom probe tomography}\vspace{.2 cm}}

\author{\large K.~Hoummada}
\affiliation{\small IM2NP, CNRS and Universit\'e Aix-Marseille, Marseille, France.}
\author{\large F.~Dahlem}
\altaffiliation{Current address: Tribology and System Dynamics Laboratory; franck.dahlem@ec-lyon.fr.}
\affiliation{\small Institut N\'eel, CNRS and Universit\'e Joseph Fourier, Grenoble, France.}
\author{\large T.~Kociniewski}
\affiliation{\small \mbox{Institut d'Electronique Fondamentale, CNRS and Universit\'e Paris Sud, Orsay, France.}}
\author{\large J.~Boulmer}
\affiliation{\small \mbox{Institut d'Electronique Fondamentale, CNRS and Universit\'e Paris Sud, Orsay, France.}}
\author{\large C.~Dubois}
\affiliation{\small \mbox{Institut des Nanotechnologies de Lyon, CNRS and INSA Lyon, Villeurbanne, France.}}
\author{\large G.~Prudon}
\affiliation{\small \mbox{Institut des Nanotechnologies de Lyon, CNRS and INSA Lyon, Villeurbanne, France.}}
\author{\large E.~Bustarret}
\affiliation{\small Institut N\'eel, CNRS and Universit\'e Joseph Fourier, Grenoble, France.}
\author{\large H.~Courtois}
\affiliation{\small Institut N\'eel, CNRS and Universit\'e Joseph Fourier, Grenoble, France.}
\author{\large D.~D\'ebarre}
\affiliation{\small \mbox{Institut d'Electronique Fondamentale, CNRS and Universit\'e Paris Sud, Orsay, France.}}
\author{\large D.~Mangelinck\vspace{.1 cm}}
\affiliation{\small IM2NP, CNRS and Universit\'e Aix-Marseille, Marseille, France.}

\begin{abstract}
Superconducting boron-doped silicon films prepared by gas immersion laser doping (GILD) technique are analyzed by atom probe tomography. The resulting three-dimensional chemical composition reveals that boron atoms are incorporated into crystalline silicon in the atomic percent concentration range,  well above their solubility limit, without creating clusters or precipitates at the atomic scale. The boron spatial distribution is found to be compatible with local density of states measurements performed by scanning tunneling spectroscopy. These results, combined with the observations of very low impurity level and of a sharp two-dimensional interface between doped and undoped regions show, that the Si:B material obtained by GILD is a well-defined random substitutional alloy endowed with promising superconducting properties.
\end{abstract}

\maketitle

In the last few years, gas immersion laser doping (GILD) technique has succeeded to provide ultra-shallow silicon junctions doped with boron atoms at a concentration well into the percent range. Samples prepared by this out-of-equilibrium technique present metallic conducting properties at room temperature~\cite{Kerrien200245} and a superconducting state below about 0.6~Kelvin~\cite{bustarret465,marcenat020501}. However, the incorporation of doping atoms well above the solubility limit is usually thermodynamically unstable, leading to phase separation and formation of aggregates~\cite{Cojocaru-Miredin_ScriptaMaterialia60}. Such material modifications would drastically affect electrical properties, thus limiting potential applications. The recent observation of a spatial dispersion in the superconducting energy gap of GILD silicon~\cite{dahlem140505} raises specifically the important question of the boron spatial distribution in these epilayers.  

In this Letter, we report on atom probe tomography (APT) measurements of the boron distribution in heavily doped silicon epilayers prepared by GILD. This accurate three-dimensional spatial study shows that the boron atoms incorporated well above their solubility limit are still randomly distributed into the silicon lattice, without forming any cluster or precipitate. In such a well-defined random substitutional alloy, the residual inhomogeneity of the boron atoms density is discussed as the origin of the spatial evolution of the local density of states.

\begin{figure}[b]
\includegraphics[width=\columnwidth]{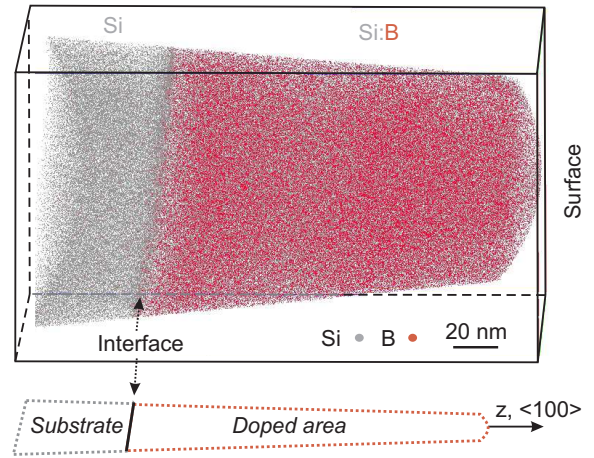}
\caption{Reconstruction of a 60x60x200~$\rm{nm^3}$ volume in heavily boron-doped silicon epilayers analyzed by atom probe tomography. Each dot corresponds to an atom. Representative of the whole distribution, only 1\% (10\%) of the captured silicon (boron) atoms are shown. A sharp interface is clearly visible between undoped and doped regions.\label{fig1_3DnB_bis2}}
\end{figure}

All silicon samples in this study were doped by GILD under the same conditions with a few $10^{21}$ boron atoms per cubic centimeter~\cite{Kerrien200245}. Starting from a $10^{-9}~\rm{mbar}$ residual pressure, a precursor gas atmosphere of boron trichloride was injected into the growth vacuum chamber to saturate the chemisorption sites of a $<$100$>$-oriented pure silicon wafer surface. This surface was melted by a homogenized 20 ns-long xenon chloride excimer laser pulse. Boron atoms then diffused into the silicon liquid, which re-crystallized from the Si/Si:B interface to the surface, in epitaxy with the substrate, see X-ray diffraction measurements in Ref.~[\onlinecite{bustarret465}] or [\onlinecite{marcenat020501}]. Such a boron injection/laser melting/re-crystallization cycle was repeated 500 times in order to heavily dope a $2 \times 2~\rm{mm^2}$ area over an expected thickness of 120~nm. The final doping level in the silicon epilayer yielded a related critical superconducting temperature of 510~mK, see transport data set B in Ref.~[\onlinecite{dahlem140505}]. 

Atom probe tomography is a 3D compositional profiling analysis based on field evaporation of atoms~\cite{kelly031101}. In the present study, GILD Si/Si:B thin films were first shaped into tips with 50~nm typical diameter using the lift-out technique~\cite{Thompson2007131} and the annular milling method~\cite{Larson1999287}. Pulsed picosecond-laser evaporation was then performed on individual tips with an Imago LEAP 3000X HR APT instrument equipped with a large-angle reflectron. The analyses were done at a specimen temperature of 50~K, a 100~kHz pulse repetition rate, an energy of 0.7~nJ (or 0.5~nJ) per pulse, an evaporation rate of 0.01 ions per pulse and a pressure of less than $10^{-11}$~mbar. In total, four tips made from two twin samples, i.e. samples prepared under the same GILD conditions, were analyzed. Similar results have been obtained for all tips. 

Fig.~\ref{fig1_3DnB_bis2} shows a reconstruction of a $60\times60\times200~\rm nm^3$ tip volume, which represents a typical result of our APT  measurements.  Because of the high efficiency of analyzed atoms by APT (around 40\% compared to few percent with secondary ion mass spectroscopy, SIMS), only a fraction of boron and silicon detected atoms is displayed in Fig.~\ref{fig1_3DnB_bis2} as red and grey dots, respectively. In this measurement, pure and doped silicon regions are easily identified from their composition. In the pure silicon region, no peak was found for boron atoms in the local mass spectrum. The detection limit was $\rm{ \sim 5.10^{18}~at./cm^3}$. The interface between Si and Si:B is well described by an isoconcentration of 2.5\% of boron atoms. It was found flat, as expected for a liquid phase epitaxy from a crystalline interface~\cite{Kerrien200245}, with a roughness of about 0.7~nm. Such a 2D sharp transition can be related to the much larger diffusion length of boron atoms in the liquid layer compared to the solid material. At the Si/Si:B junction, carbon, the dominant impurity, reaches a maximum concentration at 0.25\%, against about 0.05\% throughout the film. No chlorine atoms brought by the dopant gas~\cite{Bourguignon_SurfaceScience338} was detected in the doped layer, confirming the purity of the film.

\begin{figure}[t]
\includegraphics[width=\columnwidth]{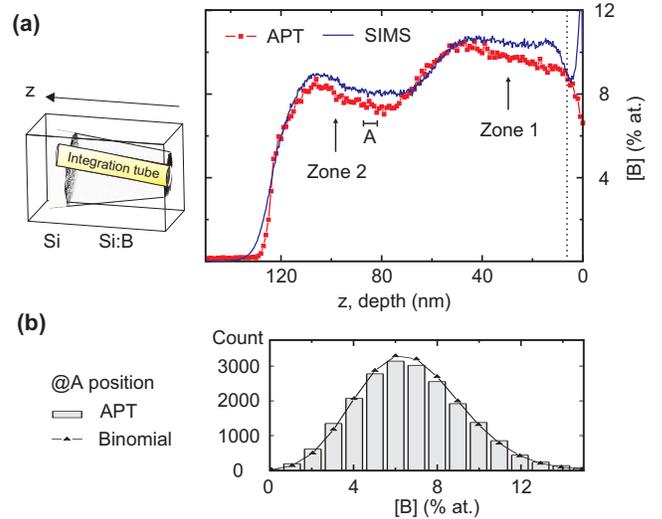}
\caption{(a) One-dimensional boron depth distribution (red square) calculated by integrating APT data of Fig.~\ref{fig1_3DnB_bis2} over a 10~nm radius cylindrical volume (see left drawing). The blue line corresponds to secondary ion mass spectroscopy measured with a CAMECA IMS4F on a twin sample. Vertical dotted line delimits the transient surface region. (b) Histogram showing the distribution of the boron contents in sets of 200 atoms located in $30 \times 30 \times 10~\rm{nm^3}$ data slice (position A) compared to a binomial shape (black line). 
\label{fig2_APTsims}}
\end{figure}

Fig.~\ref{fig1_3DnB_bis2} data provide information about Si:B region composition at the \emph{nanometer scale}, showing the important result that nanosized precipitates or clusters~\cite{Cojocaru-Miredin_ScriptaMaterialia60} of boron atoms are absent. In order to determine precisely the one-dimensional boron depth distribution, a radial integration was performed over a 10~nm diameter cylindrical region perpendicular to the Si/Si:B interface, see Fig.~\ref{fig2_APTsims}a. The related concentration profile as a function of the depth (square symbols) presents two distinct zones with an average boron atom concentration of about 10\% (zone~1) and 8\% (zone~2). These concentration levels are compatible with a boron incorporation rate estimated at 1-1.2~$\rm{10^{14}~at./cm^{2}}$ per laser shot of $\rm{1~J/cm^2}$~[\onlinecite{Kerrien_ThinSolidFilms453}]. Previous works~\cite{bustarret465,marcenat020501} showed a clear difference of behavior in terms of SIMS profile, diffraction peaks and critical temperature between samples having concentration of boron atoms above or below a value of about 4~\%.  Occuring only in the case of very heavy doping, the presence of two zones result either from a crystalline relaxation or from the lower boron atomic mobility in a highly concentrated boron silicon liquid.

High concentration SIMS profiles measured in a twin sample (full line in Fig.~\ref{fig2_APTsims}a) have been used to calibrate the z-axis of the APT reconstruction and furthermore coincide with the APT curve. The weak discrepancy is attributed to the existence of slightly different boron concentration between twin samples. Large deviations found near the surface ($\rm{ 0 \leq z \leq 5~nm}$) are well-known measurement artefacts~\cite{meuris1482}. In principle, it is difficult to quantify high concentration levels above 1 at.~\% with SIMS due to matrix effects. The present quantitative similarity between the APT and SIMS data confirms the accuracy of the recently proposed Isotopic Comparative Method used to calibrate SIMS measurements to the alloy concentration range~\cite{dubois1377}. 

Using a statistical analysis of the large volume reconstruction shown in Fig.~\ref{fig1_3DnB_bis2}, we can evaluate the dispersion in boron concentration values and determine whether or not aggregates exist at the \emph{atomic scale}. For these purposes, sets of fixed numbers of atoms have been extracted from a $30 \times 30 \times 10~\rm{nm^3}$ data slice. They have been afterward counted and classified according to their boron concentration. For instance, Fig.~\ref{fig2_APTsims}b shows the obtained distribution at the depth A for sets containing 200~atoms (reduced $\chi^2 \approx 1.1$). Several sets sizes varying from 300 to 50 atoms for one set have also been tested. The same binomial lineshape characteristic of a random distribution~\cite{Moody542} was obtained for all set sizes. By considering the smallest atomic set size, i.e. 50 atoms, and that the boron concentration in a set varies between 2 to 11\%, we conclude to the absence of boron clusters on a scale as small as 17 boron atoms (this value takes into account the detection efficiency of 37\%). Such a statement is made possible by the large measured volume, which provides from one dataset statistical information on the distribution of solute elements down to the atomic scale. Several other tests such as first nearest neighbor method~\cite{Thompson2007131}, radial distribution functions (B-B), or cluster identification routines~\cite{Heinrich_MaterialsScienceandEngineeringA353,Vaumousse_Ultramicroscopy95} were applied to different regions: no atomic cluster could be detected, either. In low boron concentration samples, Si:B composition is known to be difficult to obtain by APT due to a high evaporation field of boron compared to silicon~\cite{Ronsheim}. Here, in heavily boron-doped silicon, we might benefit from the random boron distribution.

\begin{figure}[t]
\includegraphics[width=\columnwidth]{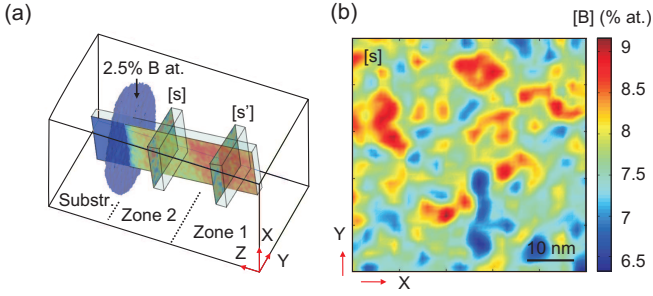}
\caption{(a) Schematic showing the position of the data boxes used to determine the lateral boron distribution. (b) Obtained distribution in a x-y plane of zone 1 at z~=~30~nm. The z integration over 10~nm was performed perpendicular to the Si/Si:B interface. The resolution is fixed by the integration over a 1~$\rm nm^2$ area. 
\label{Fig3_APT_lateral}}
\end{figure}

In comparison, for boron atoms implanted in silicon, precipitates of $\rm{SiB_3}$ stable phases form at $\rm{900^\circ\,C}$. The formation of metastable boron clusters was even observed at temperature as low as $\rm{600^\circ\,C}$. These clusters were linked to the high density of defects generated by the implantation process, such as Si self interstitials and their associated larger defects~\cite{Cojocaru-Miredin_ScriptaMaterialia60}. During GILD out-of-equilibrium doping process, the temperature is also high enough to form a $\rm{SiB_3}$ precipitate in a stable phase. It is therefore surprising that boron atoms at concentrations much higher than their solubility limit in crystalline silicon (approximately 0.2 at.\% at $\rm{1000^\circ\,C}$) do not precipitate. It is likely the high speed of recrystallization (typically 3-4~m/s)~\cite{Kerrien_ThinSolidFilms453} and the lack of implantation defects in the GILD process that hinder the precipitation of $\rm{SiB_3}$ and the related boron clustering. Kinetics overcome here thermodynamics.

From our APT measurements, we have also extracted the 2D-mapping boron atoms lateral distribution by making an integration in the z-direction from a thin box. Several z-thickness between 2 to 10~nm have been tested. The two boxes in Fig.~\ref{Fig3_APT_lateral}a represent the 10~nm case. As shown in Fig.~\ref{Fig3_APT_lateral}b, boron concentration was found to vary over a typical length of 8~nm between 6.4 and 9.1\% in zone~2, and between 7.9 and 12.1\% in zone~1. Concentrations are less spread compared to Fig.~\ref{fig2_APTsims}b since we averaged over a larger thickness. The deviation from the average value is higher for the area containing more dopants, which argues for an even better homogeneity in samples prepared with a lower boron atoms concentration. Although important modifications of the material composition like clusters and aggregates are absent in heavily-doped GILD Si epilayers, fluctuation in the distribution of uncorrelated boron atoms remains on a nanometer scale.

If these boron atoms are isolated substitutional atoms that do not form diboron or B-H pairs, such an inhomogeneous distribution of uncorrelated boron atoms is likely to be at the origin of hole density fluctuations and disorder and thus determine its relevant spatial scale. Above the superconducting transition temperature, this leads to a very dirty metal behavior~\cite{Kerrien_ThinSolidFilms453, bustarret465}. In the superconducting state, the nanometer-scale puddle-like inhomogeneities should increase localization of the superconducting Cooper pairs and thus modulate the spatial distribution of the superconductor gap. 

\begin{figure}[t]
\includegraphics[width=\columnwidth]{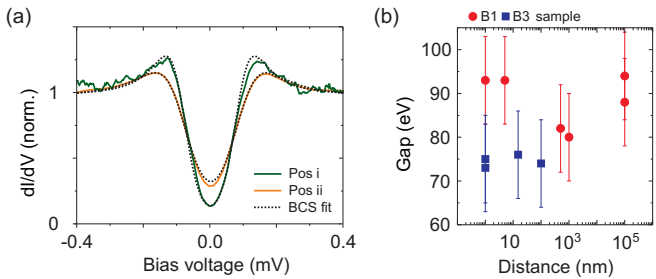}
\caption{(a) Normalized differential conductances (dI/dV) of tunnel contacts on superconducting silicon at two different spatial positions ($T_{sample}$ = 110~mK). (b) Distribution of the superconducting energy gap values obtained at different lateral distance from an arbitrary reference point on two twin samples.
\label{Fig4_STM}}
\end{figure}

Actually, superconductivity spatial fluctuations were recently observed on twin films by scanning tunneling spectroscopies~\cite{dahlem140505} in the subKelvin temperature range see Fig.~\ref{Fig4_STM}a. This detailed measurement of local electronic density of states showed that the superconductivity follow the Bardeen-Cooper-Schrieffer model and that its energy gap $\Delta (r)$ varies by up to about $\pm 15\%$ on a lateral spatial scale of about 10 nm and more. Fig.~\ref{Fig4_STM}b illustrates this inhomogeneity by presenting the energy gap value measured at different spatial positions on the surface. Every tested position showed a distinct superconductivity gap.

Our experimental observations confirm that the behaviors of the boron atoms concentration and of the superconductor energy gap are compatible both in terms of relative magnitude variation (on the order of 15\%) and characteristic length scale (of about 10 nm). Nevertheless, the observed characteristic length scale appears as significantly smaller than the calculated superconducting coherence length of about 50~nm~[\onlinecite{marcenat020501}]. The usual approach combining BCS theory and Drude model, leading to the latter value of the coherence length, thus appears not fully relevant in a very dirty metal like GILD heavily boron-doped silicon epilayers with an elastic mean free path of the order of few nanometers. 

In conclusion, boron atoms incorporated inside a silicon lattice by immersion laser doping technique do not form precipitates even when their concentration is well above the solubility limit. The doping distribution on a nanometric scale is considered responsible for the local fluctuations of the superconducting properties in Si:B. The random boron distribution combined with a good purity material and the observation of an abrupt Si/Si:B interface demonstrate the feasibility of building innovative devices based on such a GILD-doped silicon. Here, atom probe tomography appears as a perfect tool to analyse the three dimensional chemical composition of superconducting semiconductor~\cite{Blase375,Iakoubovskii2009675,bagraev21}. In the future, this technique could be used for several other covalent cubic semiconductors showing superconductvity like diamond~\cite{Ekimov_nat.428.542}, silicon carbide~\cite{ren103710} or germanium~\cite{Herrmannsdorfer217003,ren103710}.

The authors thank M.~Descoins for technical support. We acknowledge the financial support from ANR contract ANR-08-BLA-0170 and from the french CNRS (FR3507) and CEA METSA network. 


%

\end{document}